\documentclass{elsart}
\usepackage{amsmath}
\newcommand{\bm}[1]{\mbox{\boldmath$#1$}} % Remove if bm.sty is present
\newcommand{\bphi}{\bm{\phi}}
\newcommand{\bfeta}{\bm{\eta}}
\newcommand{\rmi}{\mathrm{i}}
\newcommand{\rme}{\mathrm{e}}
\newcommand{\rmd}{\mathrm{d}}
\newcommand{\eref}[1]{(\ref{#1})}
\newcommand{\opn}{\hat{\mathcal{N}}}
\newcommand{\opl}{\hat{\mathcal{L}}}
\newcommand{\mm}[1]{\mathrm{#1}}
\newcommand{\bi}[1]{\bm{#1}}
\newcommand{\mc}[1]{\mathcal{#1}}
\begin{document}
\begin{frontmatter}
\title{Statistics of noise-driven coupled nonlinear
oscillators: applications to systems with Kerr nonlinearity}

\author[FTINT]{Jaroslaw E. Prilepsky}
\ead{kovalev@ilt.kharkov.ua} and
\author[Aston]{Stanislav A. Derevyanko \thanksref{IRE}}
\ead{s.derevyanko@aston.ac.uk}
\address[FTINT]{B.I. Verkin Institute for Low Temperature Physics and Engineering,
NASU, 47 Lenin Av., 61103, Kharkov, Ukraine}
\address[Aston]{Photonics Research Group, Aston University, Birmingham, UK, B4 7ET}
\thanks[IRE]{Also at Institute for Radiophysics and Electronics, NAS of
Ukraine, 12 Acad. Proscuri St., 61085, Kharkov, Ukraine}

\begin{abstract}
We present exact analytical results for the statistics of
nonlinear coupled oscillators under the influence of additive
white noise.  We suggest a perturbative approach for analysing the
statistics of such systems under the action of a determanistic
perturbation, based on the exact expressions for probability
density functions for noise-driven oscillators. Using our
perturbation technique we show that our results can be applied to
studying the optical signal propagation in noisy fibres at
(nearly) zero dispersion as well as to weakly nonlinear lattice
models with additive noise. The approach proposed can account for
a wide spectrum of physically meaningful perturbations and is
applicable to the case of large noise strength.
\end{abstract}


\end{frontmatter}

\section{Introduction}
Weakly nonlinear coupled systems (discrete self-trapping model,
discrete nonlinear Shr\"{o}dinger equation, etc.), belong to the
universal, widely applicable, highly illustrative and thoroughly
studied models of nonlinear physics. These models have
applications to molecular crystals, molecular dynamics, nonlinear
optics, biomolecular dynamics, and so on (see e.g.
\cite{e90,Elbeck85,krb01,Elbeckonline} and references therein). It
was shown that they exhibit unusual dynamical phenomena pertaining
only to the nonlinear discrete systems, like existence of
intrinsic localised modes etc. \cite{cfk03}. Apart from the
dynamical behaviour the influence of random force and statistical
properties of nonlinear lattices are also subjects of keen
interest \cite{rcj98,mt94,rck00,f02}. The complex dynamics of such
systems is determined by an interplay between such factors as
randomness, discreetness and nonlinearity. The results concerning
the statistics of coupled nonlinear oscillators driven by Gaussian
white noise, given below, can be straightforwardly applied to the
class of problems which are naturally discrete, e.g. for the noisy
self trapping model and its modifications. The second (but not
less important) purpose of the current paper is to show that the
stochastic discrete models with finite number of degrees of
freedom can also be applied to the \textit{regularised stochastic
continuous systems} and can have a much broader field of
application. More specifically, in the spirit of the original idea
of Mecozzi \cite{Mec1,Mec2}, we show that the model of randomly
driven coupled nonlinear oscillators can be used for studying the
statistics of signal propagation in a nonlinear optical fibre with
inline noise sources.

A universal model describing the propagation of the wave envelope
in a weakly nonlinear dispersive media (which include optical
fibres) is the Nonlinear Shr\"{o}dinger Equation (NLSE) and its
scalar and vector modifications \cite{Mecozzi,Agraval,aa00}. The
need for the profound exploration of noise-stimulated fluctuations
of signal parameters in an optical fibre stems from the great
practical importance of such study for immediate technical
purposes \cite{Mecozzi,Agraval}. The pulse propagation inside an
optical fibre with inline amplifiers is described by the perturbed
NLSE with additive noise, which accounts for amplifier spontaneous
emission (ASE)\cite{Mecozzi}. It provides an illustrative example
of a complicated continuous noisy system where the exact results
concerning the statistics of the signal are still very scarce
owing to the complexity of the problem (see for instance, review
article \cite{Hausrev} and also \cite{fklt,OSA03} for recent
results). However if we assume that the noise is delta correlated
in both time and space a natural regularisation occurs, which
transforms our continuous system into a system of coupled
nonlinear oscillators subjected to an external white noise. The
statistical properties of such regularised discrete system are
much easier to derive than those of the original continuous
systems. In particular one can use the Fokker-Planck equation
(FPE) approach to obtain the evolution of the probability density
function (PDF) for all discrete components of the signal. But even
the discrete system corresponding to noisy NLSE and its
modifications is still very complicated because of the coupling
between the oscillators, which is due to the dispersion. Therefore
Mecozzi \cite{Mec1,Mec2} considered propagation of an optical
pulse affected by joint action of Kerr nonlinearity and additive
white Gaussian noise (WGN) at zero dispersion, i.e. the dispersive
(second time derivative) term in the NLSE was dropped. The
resulting effective dispersion free equation (after the
regularisation procedure) formally coincides with the dynamical
equation for a single nonlinear oscillator driven by a noisy
external force.

In the present paper we treat a zero dispersion stochastic system
(a single noise-driven nonlinear oscillator) and its
straightforward generalisations (nonlinearly coupled noisy
oscillators) as basic systems and consider various small
deterministic perturbations which can be of quite arbitrary
nature: higher order nonlinearities, small residual dispersion,
etc. In particular we can add a perturbation which will be the
direct discrete analogue of that of the corresponding continuous
optical system. Other continuous models yielding nonlinear
discrete equations addressed in the current paper, include higher
dimensional systems of nonlinearly coupled NLSEs. Two nonlinearly
coupled oscillators with particular coupling coefficients
correspond to well known Manakov equations (see e.g. \cite{lk97})
with zero dispersion driven by white noise. This continuous model
describes the nonlinear pulse propagation in a noisy birefringent
fibre at nearly zero average dispersion. Systems of more than
three coupled NLSEs are also of physical relevance \cite{kl01}. In
addition to optical communications, in the context of biophysics
the case of three nonlinearly coupled NLSEs can be applied for
studying the propagation of solitons along the three spines of an
alpha helix in protein \cite{s84}. Systems involving higher
numbers of coupled NLSEs have applications in the theory of
optical soliton wavelength division multiplexing \cite{cas95},
multichannel biparallel-wavelength optical fibre networks
\cite{yb98}, and so on. Hence we believe that the study of the
stochastic properties of such models (which, to the best of our
knowledge, is a relatively undeveloped field) is also of great
interest and may have a significant practical outcome.

As mentioned earlier, we describe the system statistics in terms
of the probability density function for the discrete variables and
the evolution of the PDF is governed by the FPE \cite{Risken}. In
the case of an unperturbed system of noise-driven nonlinear
oscillators (which is a discrete analogue of a system of zero
dispersion noisy NLSEs), the corresponding FPE can be solved
analytically. The main idea of our approach is to apply the
perturbation theory directly to the FPE rather than to the initial
stochastic system, using analytical results for the base system.
The advantage of such a method is that it allows one to obtain the
PDF for the signal output directly as a series in powers of a
small parameter $\varepsilon$, which is the effective ``strength''
of the perturbation. The perturbation theory is based on the
propagator of the unperturbed FPE. Once derived, the propagator
can be applied to various discrete systems, e.g. regularised
systems of weakly coupled NLSE at zero dispersion, where the
coupling is considered as a perturbation. This means we can
incorporate the effects of coupling and consider the statistics of
nonlinear lattice models in the so-called anticontinuum limit
(weak intersite linear coupling). In the context of optical
applications we are able to consider real polarisation mode
dispersion (PMD) systems as well as systems with wavelength
division-multiplexing (WDM) \cite{Agraval}.

The paper is organised in the following way. In section
\ref{sec:intro} we introduce basic models of one and more
noise-driven coupled nonlinear oscillators and write down the
Fokker-Planck equation for each model. After that we consider
noisy scalar and vector NLSE, describing the propagation of a
signal in the optical fibre, and show that after the
regularisation (i.e. after the introduction of a discrete time
variable) the system becomes equivalent to one of the ``base''
perturbed systems. We also mention the stochastic discrete
self-trapping model as an example of naturally discrete system
which can be analysed perturbatively.  In section \ref{sec:FP} we
proceed to find the propagators of FPEs derived for noisy
oscillator systems. Using the obtained propagators in section
\ref{sec:pertub-local} we advance to build a perturbative
expansion pf the PDF for different types of perturbations. In this
section we consider non-dispersive perturbation of one and more
coupled zero dispersion NLSE and provide explicit expressions for
several typical example systems. In section \ref{sec:nonlocal} we
move to more complicated systems, i.e. to the scalar or vector
NLSEs, where second dispersion is treated as a perturbation. In
the conclusion we summarise the results and outline the key
features and perspectives for the approach proposed.

\section{Basic models, equations and regularisation procedure} \label{sec:intro}
\thispagestyle{headings} In first two subsections we write down
the explicit form for the FPEs attached to the system of
stochastic equations, governing the evolution of oscillator
statistics. Then we explain how these systems can be employed in
the context of optical propagation and nonlinear lattices.
\subsection{Single noisy nonlinear oscillator}\label{sec:intro:1}
By adding a white noise term to the (dimensionless) dynamical
equation for a nonlinear Kerr oscillator one gains the Langevin
equation for the complex field $u= x+\rmi y$:
\begin{equation}\label{1rot}
\frac{\rmd u}{\rmd \zeta} = \mathrm{i}|u|^2u+\eta(\zeta)\, .
\end{equation}
In the mechanical interpretation $x$ is a position and $y$ is a
velocity of the nonlinear oscillator or $x$ and $y$ are the
components of a torque for the nonlinear rotator; $\zeta$ has a
meaning of time and $\eta(\zeta) = \eta_1(\zeta)+\rmi
\eta_2(\zeta)$ is the complex white Gaussian noise with the
following correlation properties:
\begin{equation}\label{1rotn}
<\eta_i(\zeta)\eta_j(\zeta')>=2D
\delta_{ij}\,\delta(\zeta-\zeta'),
\end{equation}
with $D$ being the noise strength. Equivalently one can decompose
the real and imaginary parts of \eref{1rot} and obtain a system of
two coupled Langevin equations:
\begin{equation}\label{1rotreal}
\frac{\rmd x}{\rmd \zeta} =-(x^2 + y^2) \, y + \eta_1(\zeta) \, ,
\quad \frac{\rmd y}{\rmd \zeta} = (x^2+y^2) \, x +\eta_2(\zeta) \,
.
\end{equation}
Introducing the 2D vectors $\bi{q}=\{x,y \}$ and $\bfeta = \{
\eta_1, \eta_2\}$ we can write equation \eref{1rotreal} in general
vector form:
\begin{equation}
\frac{\rmd \bi{q}}{\rmd \zeta}=\bi{f}(\bi{q})+\bfeta(\zeta).
\label{Langevin}
\end{equation}
Here $\bi{f}$ is a deterministic ``advection vector'' (note that
sometimes in the literature it is written with a minus sign). This
is the canonical form of the Langevin equation with additive WGN.
We will be interested in the statistics of the solution of
Eq.(\ref{Langevin}) for a given $\zeta$. Namely, we will seek the
conditional PDF of the vector $\bi{q}$, $P(\bi{q},\zeta)$,
provided that at the initial moment ($\zeta=0$) the PDF is given.
This function (see \cite{Risken}) obeys the following second order
partial differential equation:
\begin{equation}
\frac{\partial P}{\partial \zeta}=\sum_i \{ - \partial_i \big[ f_i
(\bi{q}) P \big] + D \, \partial^2_i  P \} \, , \label{canonical}
\end{equation}
which is a FPE attached to system (\ref{Langevin}) with the
additive white noise, given by (\ref{1rotn}). Also one has to
impose the normalisation condition $\int_{\bi q} P \, \rmd \bi{q}
=1$. A propagator is a special solution of \eref{canonical} with
the initial condition $P(\bi q, 0) = \delta(\bi q - \bi q^0)$,
i.e. it corresponds to the conditional PDF provided that at the
initial moment the system is in a deterministic state $\bi q^0$.
The explicit form of the FPE for single oscillator system
\eref{1rot} is:
\begin{equation}\label{1rotfp} \frac{\partial P}{\partial
\zeta}=D\left(\frac{\partial^2}{\partial
x^2}+\frac{\partial^2}{\partial y^2}\right) P - (x^2+y^2)\left(x
\frac{\partial}{\partial y}-y\frac{\partial}{\partial x}\right)P
\, .
\end{equation}
The boundary conditions for \eref{1rotfp} are chosen as follows:
the PDF, $P(\bi q, \zeta)$, must have no singularities and should
decrease rapidly as $|\bi q| \to \infty$ to provide the
normalisation.
\subsection{System of nonlinearly coupled oscillators}
\label{sec:intro:2} Next we study the following system of $M$
coupled nonlinear oscillators:
\begin{eqnarray}
\frac{\rmd u_1}{\rmd \zeta}& =& \rmi u_1 \sum_{i=1}^M a_{1\,i}
|u_i|^2 + \eta^1(\zeta)\nonumber \\
\vdots &&  \label{Krot} \\
\frac{\rmd u_M}{\rmd \zeta} & = & \rmi u_M \sum_{i=1}^M a_{Mi}
|u_i|^2 + \eta^M(\zeta) \, . \nonumber
\end{eqnarray}
Here again $u_i = x_i + \rmi y_i$, are the complex fields,
$a_{ij}$ are real constant coefficients  and the correlation
properties for complex WGNs $\eta^k=\eta^k_1+\rmi \eta^k_2$ are
(here and further on the subindexes of complex WGNs, ``1" and
``2", imply the real and imaginary parts correspondingly):
\begin{equation}\label{2rotn}
<\eta^k_i(\zeta)\eta^l_j(\zeta')>=2D_k \delta_{kl}
\delta_{ij}\,\delta(\zeta-\zeta'),
\end{equation}
(here no summation over the repeated index $k$ implied). To be
more general we suppose that WGNs $\eta^k(\zeta)$ may have
(possibly) different intensities $D_k$. Again we can introduce
2M-dimensional real vectors $\bi q=\{x_1,\ldots,x_M,y_1,\ldots
y_M\}$ and $\bfeta=\{\eta_1^1,\ldots \eta_1^M, \eta_2^1, \ldots
\eta_2^M\}$, and after singling out real and imaginary parts in
\eref{Krot} arrive at the generic Langevin equation
\eref{Langevin}. The FPE for this system has the form
\begin{equation}\label{2rotfp}  \frac{\partial P}{\partial
\zeta}= \sum_{i=1}^M \left( D_i \Delta_i  - \sum_{j=1}^M
a_{i\,j}(x_j^2+y_j^2) \left[ x_i \frac{\partial}{\partial
y_i}-y_i\frac{\partial}{\partial x_i} \right] \right)P \, .
\end{equation}
Here $\Delta_i = \partial^2/\partial x_i^2 + \partial^2/\partial
y_i^2$. We shall call the number $M$ of nonlinearly coupled
equations in (\ref{Krot}) the \textit{dimensionality} of the
system (the number of independent variables in the FPE
(\ref{2rotfp}) is then $2M$).
\subsection{Regularisation procedure for optical pulse propagation:
discrete time}\label{sec:intro:3} The propagation of a complex
light envelope $u(\zeta,t)$ in a noisy optical fibre line with
so-called weak dispersion management (see Refs.
\cite{Mecozzi,Tur00}) is described by the stochastic NLSE:
\begin{equation}
\frac{\partial u}{\partial
\zeta}=-\mathrm{i}\,\frac{\bar{\beta}}{2}\, \frac{\partial^2
u}{\partial t^2}+ \mathrm{i}|u|^2u+\eta (t,\zeta).\label{nlse}
\end{equation}
(Note that here $\zeta$ plays the role of coordinate in contrast
to the mechanical interpretation of oscillators and ``virtual''
time is $t\,$.) In equation \eref{nlse} the stochastic term
$\eta(t,\zeta)$ represents a complex WGN. The following
correlation properties are assumed:
\begin{equation}
\begin{split}
&<\eta(t,\zeta)>= <\eta(t,\zeta)\eta(t',\zeta')> =0, \\
&<\eta^*(t,\zeta)\eta(t',\zeta')>= 2 G \delta(\zeta-\zeta')
\delta(t-t'). \label{c1}
\end{split}
\end{equation}
Here $G$ is a normalised noise strength. Equation (\ref{nlse})
describes the path-averaged model of the dispersion managed
optical fibre communication system and, it was shown (see
\cite{Tur00} and references therein for details) that this model
is applicable when the so-called strength of the map, proportional
to the local fibre dispersion, to the dispersion oscillation
period and to the squared reciprocal pulse width, is small. This
strength parameter characterises the effect of the variation of
local dispersion. Parameter $\bar{\beta}$ entering Eq.(\ref{nlse})
is just a path-averaged dispersion coefficient. A zero dispersion
NLSE occurs when the path-averaged dispersion coefficient is equal
to zero, $\bar{\beta}=0$. The statistics of system (\ref{nlse}) at
zero dispersion point were first studied in \cite{Mec1,Mec2} (see
also \cite{PRL03}). In these Refs. an analytical formula for the
conditional PDF of the output signal $u(t,L)$ given the
deterministic input signal $u_0(t,0)$ was obtained.

Here we would like to consider the zero-dispersion NLSE assuming
also that there is an addition corresponding to a deterministic
perturbation:
\begin{equation}\label{nl}
\frac{\partial u}{\partial \zeta}=\mathrm{i}|u|^2u+\eta(t,\zeta) +
\varepsilon \hat N[u, \ldots] \, .
\end{equation}
Here $\varepsilon \ll 1$ is intended and let $\hat N[...]$ be any
sort of perturbation of the initial zero-dispersion NLSE: it may
be either a ``nonlocal'' operator involving the time
$\partial/\partial t$ derivatives $\hat N[u,u'_t,u''_{tt},
\ldots]$, or some kind of a ``local'' operator, $\hat N[u]$. Note
that complex WGN defined by expression (\ref{c1}) is equivalent to
the two independent real WGN,
$\eta(t,\zeta)=\eta_1(t,\zeta)+\mathrm{i}\eta_2(t,\zeta)$, with
\begin{equation}
\begin{split}
<\eta_1(t,\zeta)>=&<\eta_2(t,\zeta)>=<\eta_1(t,\zeta)\eta_2(t',\zeta')>=0
\, ,
\\ <\eta_1(t,\zeta)\eta_1(t',\zeta')> = &<\eta_2(t,\zeta)\eta_2(t',\zeta')>=
G \delta(\zeta-\zeta') \delta(t-t') \, . \label{mean}
\end{split}
\end{equation}
Because the white noise is delta correlated in both time and
space, it has an infinite average power $<|\eta(t,\zeta)|^2>$.
This is the consequence of the idealised character of the white
noise. In practice, real noise always has small but finite
correlation length $r_c$ and correlation time $\tau_c$. For ASE in
optical fibre transmission links the correlation length is in the
order of the average amplifier spacing $L_a$, and correlation time
is inversely proportional to the bandwidth of the noise: $\tau_c
\sim B^{-1}$. In what follows we will consider our noise still
delta correlated in space. However to make the correspondence
between discrete and continuous systems explicit and also to get
physically meaningful results we move to the context of discrete
signals and introduce discrete time variable $t$. We will consider
the case where both signal and noise are entirely contained to the
time interval of length $T$. Then, to sample the signal
effectively, we need a finite number of discrete samples,
separated by an interval $\Delta t \sim \tau_c \ll T$. If we put
$\Delta t = B^{-1}$, the number of samples is $N=BT$. The signal
is represented as $N$-dimensional complex vector,
$\bi{u}=\bi{x}+\mathrm{i}\bi{y}$, in the sample space:
\begin{equation}
u(t,\zeta) \Rightarrow \bi{u}(\zeta), \quad u_i(\zeta)=u(i\Delta
t,\zeta), \quad i=1,\ldots N \, . \label{regul}
\end{equation}
Analogously one can work in the real $2N$-dimensional space of
real and imaginary parts of vector $\bi{u}$:
$\bi{q}=\{x_1,\ldots,x_N,y_1,\ldots,y_N\}$. After the
regularisation Eq.(\ref{nl}) is equivalent to the system of
equations:
\begin{equation}\label{nl1}
\frac{\rmd u_i}{\rmd \zeta}=\mathrm{i}|u_i|^2u_i+\eta^i(\zeta) +
\varepsilon \hat N_i[\bi{u}] \, ,
\end{equation}
Vector
$\bi{\eta}(\zeta)=\bi{\eta}_1(\zeta)+\mathrm{i}\bi{\eta}_2(\zeta)$
is a regularised white noise with the same properties as given by
(\ref{mean}) but with regularised temporal delta functions:
\begin{equation}\label{cor-r}
<\eta^i_{1}(\zeta)\eta^j_{1}(\zeta')>=
<\eta^i_{2}(\zeta)\eta^j_{2}(\zeta')>= 2D
\delta_{ij}\,\delta(\zeta-\zeta'),
\end{equation}
where we set $D \equiv BG$. The operator $\hat{N}_i$ is
discretised in time as well.

We call the equation, governing the unperturbed stochastic
evolution of each separate sampled component, the \textit{base
system}. For all cases considered in this paper, a base system
will be either of the form Eq.(\ref{1rot}) or Eq.(\ref{Krot}); the
corresponding FPE for each base system has either a form
(\ref{1rotfp}) or a more general look (\ref{2rotfp}), depending on
the dimensionality of the base system.

One can see that each sampling component in Eq.(\ref{nl1}) obeys
the equation governing the noisy dynamics of a single nonlinear
oscillator, Eq.(\ref{1rot}): hence the dimensionality of the
corresponding base system is one. The complete set of equations
for all components of the sampled signal has a form of perturbed
Eq.(\ref{Krot}) with $M=N$ and an identity matrix of
$a$-coefficients: $a_{ij}=\delta_{ij}$. As was noted before, we
divide the perturbations in equation (\ref{nl1}) into two
categories: ``local'', which depend on the value of the signal at
the current moment of time only, i.e.
$\hat{N}_i[\bi{u}]=\hat{N}_i[u_i]$, and ``nonlocal'' (as, for
instance, dispersion, involving second or higher time derivatives
of the signal), which comprise the field values at other sampling
points.

The same regularisation procedure can be performed with other
models relevant to the noisy signal propagation. For example we
can consider the propagation in the birefringent fibres where
instead of NLSE, under certain conditions, we have a set of
 Manakov equations \cite{lk97} perturbed by noise:
\begin{equation}\label{man}
\begin{split}
 \frac{\partial u}{\partial
\zeta} &= -\rmi \frac{\bar \beta}{2} \frac{\partial^2 u}{\partial
t^2} + \rmi u \left( |u|^2 + |v|^2\right) +
\eta^1(\zeta,t) \, ,\\
\frac{\partial v}{\partial \zeta} &= -\rmi \frac{\bar \beta}{2}
\frac{\partial^2 u}{\partial t^2} +  \rmi  v \left( |u|^2 +
|v|^2\right) + \eta^2(\zeta,t) \, .
 \end{split}
\end{equation}
Here $u$ and $v$ stand for complex components of the left and
right polarised waves and $\eta^i(\zeta,t)$ are the complex WGNs.
Without noise addition this system describes the processes of the
two-dimensional stationary self-focusing and one-dimensional
auto-modulation of electromagnetic waves with arbitrary
polarisation. Neglecting the dispersion and adding perturbation,
after the regularisation one obviously gains:
\begin{equation}\label{man1}
\begin{split} \frac{\rmd u_i}{\rmd \zeta}&=\mathrm{i}(|u_i|^2+|v_i|^2)u_i +
\eta^1_i(\zeta) + \varepsilon \hat
N^1_i[\bi{u},\bi{v}] \, , \\
\frac{\rmd v_i}{\rmd \zeta}&=\mathrm{i}(|u_i|^2 + |v_i|^2) v_i +
\eta^2_i(\zeta) + \varepsilon \hat N^2_i[\bi{u}, \bi v ] \, .
\end{split}
\end{equation}
In this case, for each separate sampled pair $[u_i,v_i]$ (if
$\varepsilon$ is again set to zero), we arrive at the system of
two nonlinearly coupled oscillators. This is just a special case
of Eq.(\ref{Krot}), where the number of dimensions of the base
system is two. The whole system for $K=2N$ sampled components of
the signal ($N$ is, as previously, the number of the sampling
time-points) has the form of Eq.(\ref{Krot}) as well. However, as
the base system has a dimensionality two, here the coupling matrix
$a_{ij}$ is no longer an identity matrix but have a $(2 \times
2)$-block diagonal form. The ``local'' perturbations for such
systems are those which satisfy: $\hat N_i[\bi u , \bi v] = \hat
N_i[u_i,v_i]$.

In the same way one could deal with the systems involving higher
number $M$ of coupled NLSEs: the number of equations would simply
define the dimensionality $M$ of the corresponding base system.
The regularised system for the whole set of time-sampled signal
components would possess the matrix $a_{ij}$ in a block-diagonal
form, where each block would be an $(M \times M)$ square matrix.

\subsection{Nonlinear lattice model}\label{sec:intro:4}
The dynamics of a weakly nonlinear lattice is described by the set
of differential-difference equations (the discrete self-trapping
equation (DSTE), see Refs.\cite{e90,Elbeck85,Elbeckonline}):
\begin{equation}\label{dste}
\frac{\rmd u_i}{\rmd \zeta} = \mathrm{i}|u_i|^2u_i + \rmi \,
\varepsilon
 \sum_{j} m_{ij} u_j +\eta^i(\zeta)\, ,
\end{equation}
where we have also added the white noise terms $\eta^i(\zeta)$ at
each site (with the same statistical properties as in
\eref{cor-r}). In Eq.\eref{dste} $m_{ij}$ is a coupling matrix and
index $j$ ranges over the whole 1D lattice containing $N$ sites.
The generalisation for lattices of higher dimensions is
straightforward. In physical applications $m_{ij}$ is real and
symmetric; in the case of tridiagonal matrix, i.e. for the nearest
neighbor coupling, we recover a particular realisation of DSTE --
the discrete nonlinear Shr\"{o}dinger equation \cite{krb01}.

Equation (\ref{dste}) provides an example of a naturally discrete
system with Kerr nonlinearity. If we consider the case of
vanishing $\varepsilon$, the so-called anticontinuum limit, we
arrive at the system of uncoupled oscillators, which is obviously
the unperturbed system \eref{nl1}. Here the dimensionality of the
base system is one. The perturbation in Eq.\eref{dste} provides an
example of a nonlocal perturbation, since it involves other
lattice sites.

\subsection{The list of the notations}\label{sec:intro:5}
In this subsection we summarise the notations that will be used
throughout the paper. The aim of the proposed notations and
indexing scheme is to demonstrate the general ideology and to
underline the similarity of all final expressions: very different
physical systems after the regularisation yield similar formal
expansions for the PDF.

\begin{table}[h!]
\caption{Table of notations and indexing scheme used in the paper}
\begin{tabular}{| p{3.5cm} | p{9.7cm} |}
\hline \begin{center} \textbf{Notation} \end{center} & \begin{center} \textbf{Meaning} \end{center} \\
\hline $\bi u$, $\bi v$, $u_i$, $v_i$ & The sampled complex field
variables. The bold symbols indicate the whole sets $\bi u =
\{u_1, u_2, \ldots \}$ etc. The subscript indicates the sampling
time-point number.\\
\hline $x_i$, $y_i$ and $r_i$, $\phi_i$ & Real and imaginary parts
and modulus-phase representation of the complex field variables.
The subscript shows the sampling number (in some examples the
upper index will be used as well).
\\ \hline
$\bi r$, $\bphi$, $\bi q$, $\tilde{\bi{q}}$  & Bold symbols denote
sets $\bi r \equiv \{r_1, r_2, \ldots \}$, $ \bphi \equiv
\{\phi_1, \phi_2,\ldots \}$; where indicated explicitly, the
subindex shows the dimensionality of the set. Vectors $\bi q$ and
$\tilde{\bi q}$ are the joint sets: $\bi q_K \equiv \{x_1, x_2,
\ldots, x_K, y_1, y_2, \ldots, y_K \}$, and $\tilde{\bi q}_K
\equiv \{r_1, r_2, \ldots, r_K, \phi_1, \phi_2, \ldots, \phi_K
\}$. The superscript ``0'' marks the initial deterministic values.
The upper index is also used to distinguish between the different
sets.
\\ \hline
$\opl_K$ &  The coordinate part of the FP-operator. The subscript
shows the (half)number of independent variables involved.\\
\hline $P_M(\bi q; \zeta)$, $P^K(\bi q; \zeta)$, $\mc P^K(\bi q;
\zeta)$ & $P_M(\bi q; \zeta)$, $P^K(\bi q; \zeta)$ are the PDFs
for the base system and for the whole sampled signal. For
convenience with a subscript we shall explicitly mark the
dimensionality of the underlying base system $M$, i.e. for $P_M$
we have $\bi q = \bi{q}_M$. A superscript will be used to mark the
whole PDF, and to show the total (half)number of the independent
variables involved. If the dimensionality of the base system is
$M$, and the number of sampling points is $N$, then $K=M \times
N$, $\bi q = \bi q_K$. The calligraphic $\mc P^K$ will mark the
PDFs of the perturbed systems. The tilde above the PDF will mean
the representation in ``polar coordinates'', $\{ \bi x , \bi y \}
\to \{\bi r , \bphi \}$.\\ \hline
\end{tabular}
\end{table}
\newpage
\begin{tabular}{| p{3.5cm} | p{9.7cm} |}
\multicolumn{2}{l}{Table of notations and designations used in the
paper (cont'd)}\\ \hline $\hat N_i[\ldots]$, $\hat{\mc N}[\ldots]$
& $\hat N_i[\ldots]$ and $\hat{\mc N}[\ldots]$ are the regularised
perturbation operator and its FP-image correspondingly. The
explicit form of the FP-image is derived using the general rules
for the construction of the FPE \cite{Risken}.
\\ \hline
$\bfeta = \{ \eta^1,\eta^2, \ldots \}$ & Complex regularised WGN.
The superscript is used to distinguish between the different
independent WGNs entering different equations.
\\ \hline
$\mc G_M(\ldots)$, $\bi{\mc G}_K(\ldots)$ & $\mc G_M(\ldots)$ is
the propagator of the base system FPE with the coordinate part
$\mc{L}_M$. Symbol $\bi{\mc G}_K$ will mark the propagator of the
whole FPE of the entire sampled system.
\\ \hline
\end{tabular}

\section{Propagator functions of the unperturbed Fokker-Plank equation}
\label{sec:FP} We have just shown that after the regularisation
all the models considered in the previous section yield same type
of system of Langevin equations, which is merely a perturbed
system of coupled nonlinear oscillators (\ref{Krot}). Now consider
the statistical properties of unperturbed system (\ref{Krot}). The
FPE attached to system \eref{Krot} is of course Eq.\eref{2rotfp}.
In this section we will calculate a propagator for
Eq.\eref{2rotfp} which will allow us to build an effective
perturbative expansion for the PDF in the next sections.

First we note that in some examples (like for instance a scalar
regularised zero dispersion NLSE \eref{nl1} with $\varepsilon=0$)
the coupling matrix $a_{ij}$ is diagonal, and the dimensionality
of the base system is one. In that case FPE \eref{2rotfp} admits
factorised solutions in the form:
\begin{equation}
P^K(\bi{q}_K;\zeta)=P(x_1,\ldots,x_K,y_1,\ldots y_K;\zeta)=
\prod_{i=1}^{K} P_1(x_i,y_i;\zeta) \, , \label{factor}
\end{equation}
where $P_1(x,y;\zeta)$ is a PDF which satisfies unperturbed FPE
for a single oscillator \eref{1rotfp}. The factorised form of
Eq.(\ref{factor}) is a natural manifestation of the fact that for
diagonal $a_{ij}$ all nonlinear oscillators in system \eref{Krot}
become dynamically uncoupled and, because of \eref{2rotn},
statistically uncorrelated. Therefore in order to study the
statistics of such a system it is sufficient to study the
statistics of a single oscillator \eref{1rot}. Bearing this in
mind we initially consider a FPE for a single oscillator
\eref{1rotfp} and find its propagator, and then utilise the
obtained results to find the propagator of generic system
\eref{2rotfp} by virtue of Eq.(\ref{factor}).

The same is true for the unperturbed systems for which a base
system has a higher number of dimensions, say $M$. In such a case
the coupling matrix $a_{ij}$ is $(M \times M)$-block-diagonal, and
the expression for the total PDF has again a factorised form:
\begin{equation}
P^K(\bi{q}_K;\zeta)=P(x_1,\ldots,x_K,y_1,\ldots y_K;\zeta)=
\prod_{i=1}^{N} P_M( \bi{q}_M^i;\zeta) \, , \label{Kfactor}
\end{equation}
where $ P_M( \bi{q}_M^i;\zeta)$ is now a PDF, which satisfies
unperturbed FPE for a system of nonlinearly coupled oscillators
\eref{2rotfp} and $N$ is the number of samples so that $K=N\times
M$. The factorisation in Eq.(\ref{Kfactor}) is possible because
each of the $N$ block-systems of $M$ nonlinear oscillators is
uncoupled from all other remaining block-systems. Now to gain the
statistics of this system we are to study the statistics of
coupled oscillators, described by FPE \eref{2rotfp}. After it is
done we employ  Eq.\eref{Kfactor} to find the final answer for the
statistics of the entire set of samples.
\subsection{Propagator of the Fokker-Plank operator for a single oscillator}
Let us study the equation for a single oscillator \eref{1rotfp}.
Proceeding to the polar coordinates $x=r\cos \phi$, $y=r\sin \phi$
we introduce function $\tilde{P}_1(r,\phi;\zeta) =
P_1(x(r,\phi),y(r,\phi);\zeta)$. Note that function $\tilde{P}_1$
\textit{is not} the PDF in polar coordinates since it does not
have the Jacobian $r$ included. For the function $\tilde{P}_1$ we
now obtain the following equation:
\begin{equation}
 \frac{\partial \tilde{P}_1}{\partial \zeta} +
\opl_1 \tilde{P}_1 = \frac{\partial \tilde{P}_1}{\partial \zeta} +
\left[- D \Delta^r  + r^2 \frac{\partial}{\partial \phi}
\right]\tilde{P}_1 \, =0 , \label{FP1}
\end{equation}
where
\[ \Delta^r =\frac{1}{r} \frac{\partial}{\partial r} \left(
r \, \frac{\partial}{\partial r} \right) + \frac{1}{r^2}
\frac{\partial^2}{\partial \phi^2} \, ,
\]
is the Laplace operator in polar coordinates. We define a
\textit{propagator}, $\mathcal{G}_1$, as a special solution of
(\ref{FP1}) (as a function of $\{\zeta, r, \phi \}$), possessing
the following properties:
\begin{equation}\label{ret2}
\begin{split}
\mathcal{G}_1(\zeta,\zeta';\phi, \phi';r,r') \Big|_{\zeta=\zeta'}
&= \delta(\phi-\phi')
\delta(r-r')/r \, ,\\
\mathcal{G}_1(\zeta,\zeta';\phi,\phi';r,r') & =0, \quad
\zeta<\zeta' \, .
\end{split}
\end{equation}
Note that propagator \eref{ret2} coincides with the retarded Green
function of Eq. \eref{FP1}.  Since our system is uniform in
$\zeta$ and $\phi$, the propagator depends only on the differences
$\Delta\zeta=\zeta-\zeta'$ and $ \Delta \phi =\phi - \phi'$:
$\mathcal{G}_1=\mathcal{G}_1(\Delta \zeta;\Delta \phi;r,r')$. To
find the propagator we first determine eigenfunctions and
eigenvalues of the Fokker-Planck operator $\opl_1$ in \eref{FP1}.
The right, $\Psi$, and left, $\tilde \Psi$, eigenfunctions are
calculated in the Appendix \ref{sec:appendix}. Using these
eigenfunctions \eref{eigensystem1}, \eref{eigensystem2}, we can
write the sought propagator as an eigenvalue expansion:
\[
\mathcal{G}_1(\Delta \zeta;\Delta \phi;r,r')=\sum_{n,\,\nu}
\Psi_{n\nu}(r,\phi)\,\mathrm{e}^{-s_{n\nu}\Delta \zeta}
\tilde\Psi^*_{n\nu}(r',\phi').
\]
Using identity \eref{ident1} the sum over $n$ can be convoluted
yielding the following result:
\begin{multline}
\mc{G}_1(\Delta\zeta;\phi - \phi';r,r') = \frac{1}{2\pi D}
\sum_{\nu = - \infty}^{\infty} \mathrm e^{\mathrm{i}\nu(\phi -
\phi')}\, \frac{\alpha_\nu}{\sinh (2\alpha_\nu \Delta \zeta)}
\\
\times\exp \left[-\frac{\alpha_\nu(r^2+r'^2)}{2D}\, \coth
(2\alpha_\nu \Delta \zeta)\right]
\mm{I}_{|\nu|}\left(\frac{\alpha_\nu r r'}{D\sinh (2\alpha_\nu
\Delta
 \zeta)}\right). \label{green}
\end{multline}
Here $\mm{I}_{|\nu|}$ stands for the $\nu$th order modified Bessel
function of the first kind and the quantities $\alpha_{\nu}$ are
defined in the Appendix \ref{sec:appendix}. Formula \eref{green}
is valid for $\Delta \zeta \geq 0$. For $\Delta \zeta <0$ the
propagator, $\mathcal{G}_1$, must be equal to zero due to
\eref{ret2}. Note that the obtained propagator is symmetric with
respect to variables $(r,r')$ and is normalised to unity:
$\int^{\infty}_0 \rmd r r \int^{2\pi}_0 \rmd \phi \,
\mathcal{G}_1(\Delta\zeta;\phi - \phi';r,r')=1$.

Propagator $P_1=\mathcal{G}_1(\zeta-\zeta';\phi-\phi';r,r')$ (with
$r(x,y)=\sqrt{x^2+y^2}$, $\phi(x,y)=\arctan (y/x)$ and the same
for primed variables) represents the solution of equation
\eref{1rotfp} provided that at the point $\zeta=\zeta'$ our
variables have deterministic values: $x=x', y=y'$. If we want to
apply our results to the regularised scalar NLSE with additive
white noise, we must use formula (\ref{factor}). Function $P_1$
describing the statistics of the unperturbed NLSE at zero
dispersion coincides with that obtained in \cite{Mec2} with a
different approach (see also Ref. \cite{PRL03}).
\subsection{Propagator for the Fokker-Planck operator for
$M$ nonlinearly coupled oscillators} In the same way, generalising
the derivation of PDF for a single oscillator, we can obtain the
PDF $\tilde{P}_M(\bi r, \bphi, \bi r_0, \bphi_{0}; \zeta)$ for
system of $M$ nonlinearly coupled oscillators (see Appendix
\ref{sec:appendix}). FPE \eref{2rotfp} in polar coordinates reads
as:
\begin{equation}
\frac{\partial \tilde{P}_M}{\partial \zeta} + \opl_M \tilde{P}_M
=\frac{\partial \tilde{P}_M}{\partial \zeta} +\sum_{i}^M \left[-
D_i \Delta^r_i + \sum_{j}^M  r^2_i
a_{j\,i}\frac{\partial}{\partial \phi_j} \right]\tilde{P}_M \, =0,
\label{FP1N}
\end{equation}
where $\Delta^r_i$ denotes the Laplacian in polar coordinates
$\{r_i,\phi_i\}$. Similarly to the case of one oscillator, we
introduce propagator $\mathcal{G}_M\,(\Delta\zeta; \bphi - \bphi';
\bi r, \bi r')$ for Eq. \eref{FP1N}, which is a solution of
\eref{FP1N} for $\Delta \zeta >0$ subject to conditions:
\begin{equation}\label{ret3}
\begin{split}
\mathcal{G}_M(\Delta \zeta;\bphi-\bphi';\bi r, \bi r')
\Big|_{\Delta \zeta=0} &=\prod_{i=1}^M \delta(\phi_i-\phi_i')
\delta(r_i-r'_i)/r_i \, , \\
\mathcal{G}_M(\Delta \zeta;\bphi-\bphi';\bi r, \bi r') & =0, \quad
\Delta \zeta< 0 \, .
\end{split}
\end{equation}
Again the propagator coincides with the retarded Green function
for Eq.\eref{FP1N}. Using eigenfunctions \eref{eigensystem34} and
convolution formula \eref{ident1} we obtain the following
expression:
\begin{equation}\label{greenN}
\begin{split}
&\mathcal{G}_M\,(\Delta\zeta; \bphi - \bphi'; \bi r, \bi r') =
\frac{1}{(2 \pi )^M} \sum_{\nu_1= - \infty}^{\infty} \ldots
\sum_{\nu_M= - \infty}^{\infty} \prod_{i=1}^M \frac{1}{ D_i}\,
\mathrm e^{\mathrm{i}\nu_i(\phi_i - \phi'_i)}\,
\\ & \times \frac{\alpha_i}{\sinh (2\alpha_i \Delta \zeta)}\,  \exp
\left[-\frac{\alpha_i(r_i^2+r^{\prime 2}_i)}{2D_i}\, \coth
(2\alpha_i \Delta \zeta)\right]
\mm{I}_{|\nu_i|}\left(\frac{\alpha_i r_i r'_i}{D_i\sinh (2
\alpha_i \Delta
 \zeta)}\right),
\end{split}
\end{equation}
where quantities $\alpha_i$ are defined in Appendix
\ref{sec:appendix}, see Eq.\eref{cases}. Eq.\eref{greenN} is one
of the major results of the current paper: it generalises the
results obtained by Mecozzi \cite{Mec1,Mec2} and Turitsin et al.
\cite{PRL03} for a single oscillator. As mentioned earlier, if the
matrix coefficients $a_{ij}$ are diagonal, then each coefficient
$\alpha_i$ in \eref{greenN} depends on $\nu_i$ only, and
propagator \eref{greenN} factorises into the product of $M$
identical functions, each having the form of Eq.\eref{green}.
However, if the coupling matrix $a_{ij}$ is nondiagonal (the case
of nonlinearly coupled oscillators), formula \eref{greenN} cannot
be reduced to a trivial product of identical components.

\section{Perturbational approach for zero-dispersion NLSE}
\label{sec:pertub-local}
\subsection{General remarks}\label{subsec:pert}
We are now in a position to build a general perturbation theory
for the perturbed zero-dispersion NLSE and its generalisations. In
this section we will consider perturbations $\hat{N}$ which do not
contain a dispersive term. Any perturbation operator $\hat N$ in
Eq.\eref{nl} after regularisation yields a Fokker-Plank image
$\opn$ which will appear in the r.h.s. of Eq.\eref{FP1N}, and will
retain a small perturbation parameter $\varepsilon$. Instead of
equation \eref{FP1N} we arrive at the perturbed equation:
\begin{equation}
\frac{\partial \tilde{\mc P}_K}{\partial \zeta} + \opl_K
\tilde{\mc P}_K = \varepsilon \opn \tilde{\mc P}_K \, ,
\label{FP1N-pert}
\end{equation}
where $K=M \times N$, $M$ is the dimensionality of the base system
and $N$ is the number of samples. Since the coupling matrix
$a_{ij}$ is block-diagonal, we have $\opl_K = \sum_{i=1}^{N}
\opl_M^i$. When we consider local perturbations to system
\eref{nl} (or regularised system \eref{nl1}), the FP-image of the
perturbation also decomposes into a sum of independent operators,
$\opn=\sum_{i=1}^N \opn_i \label{N-sum}$, each acting on a
specific set of variables $\{ \bi r^i, \bphi^i \}$ belonging to a
separate block. Using propagator (\ref{greenN}) we arrive at the
following expansion (we write only the first corrections to the
unperturbed propagator):
\begin{equation}\label{1n1}
\tilde{\mc P}^K(\tilde{\bi q} |\, \tilde{\bi q}^0;\zeta) = \tilde
P^K(\tilde{\bi q} |\, \tilde{\bi q}^0;\zeta) + \varepsilon \tilde
P^K_{(1)}(\tilde{\bi q} |\, \tilde{\bi q}^0;\zeta) + \mc O
(\varepsilon^2) \, ,
\end{equation}
where for $\tilde P^K_{(1)}$ we have:
\begin{eqnarray}
\tilde P^K_{(1)}(\tilde{\bi q} |\, \tilde{\bi q}^0;\zeta) =
\intop_0^{\zeta} &\rmd& \zeta' \intop_{Q} \rmd \tilde{\bi q}_K' \,
\bi{\mc{G}}_K(\zeta - \zeta';
\ \bphi_K- \bphi'_K;\bi r_K, \bi r'_K) \nonumber \\
& \times & \opn^{\,'} \bigl[ \bi{\mc{G}}_K(\zeta'; \ \bphi'_K-
\bphi^0_K;\bi r^0_K, \bi r'_K) \bigr], \label{+psi12}
\end{eqnarray}
\begin{equation}
\label{greenpert} \bi{\mc{G}}_K(\zeta - \zeta'; \ \bphi_K -
\bphi'_K; \bi r_K , \bi r'_K) = \prod_{i=1}^N \mc G_{M}(\zeta -
\zeta'; \ \bphi_M^i- \bphi_M^{i \, '};\bi r_M^i, \bi r^{i \, '}_M)
\, ,
\end{equation}
\begin{equation}
\intop_Q  \rmd  \tilde{\bi q}'_K \equiv
\underbrace{\intop_{0}^{\infty} \rmd r_1'\,r_1' \intop_{0}^{2 \pi}
\rmd \phi_1' \, \cdots \, \intop_{0}^{\infty}   \rmd r_K'\,r_K'
\intop_{0}^{2 \pi} \rmd \phi_K'}_K \, .
\end{equation}
The prime in $\opn^{\,'}$ indicates that operator $\opn$ acts on
the primed variables. Note that since perturbation enters only
advection terms in Langevin equation (\ref{Langevin}) (or
(\ref{nl1}))  its Fokker-Planck image has a form of divergence:
$\opn=\vec{\nabla} \cdot (...)$ (see equation (\ref{canonical})).
Recalling also that the propagator $\mathcal{G}_M$ is normalised
we can verify from equations (\ref{1n1})-(\ref{+psi12}) that the
overall normalisation of the function $\tilde{\mc P}_K$ is
preserved since $\int_Q \rmd \tilde{\bi q}'_K \tilde
P^K_{(1)}(\tilde{\bi q} |\, \tilde{\bi q}';\zeta) =0$. The PDF in
cartesian coordinates $\mc P^K (\bi q|\bi q^0;\zeta)$ can be
restored via the substitution
\[
\mc P^K (\bi q|\bi q^0;\zeta)=\tilde{\mc P}^K( \bi r(\bi x,\bi y),
\bphi(\bi x,\bi y)|\, \bi r^0(\bi x^0, \bi y^0),\bphi^0(\bi x^0,
\bi y^0);\zeta)
\]
In a similar fashion, continuing this recurrent procedure we can
find the higher corrections to the PDF \eref{1n1}.

Let us now consider explicitly some examples of nondispersive
perturbations.

\subsection{High order nonlinearities} First we consider the easiest case where the
dimensionality of the corresponding base system is one and a
perturbation is local. It seems quite natural to choose the higher
nonlinear terms as a perturbation for the single zero-dispersion
noisy NLSE (\ref{nl}):
\begin{equation}\label{n1}
\hat N [u] = \mathrm{i}\sum_{k = 1}^{q} c_k \, u |u|^{2 k+2} \, ,
\end{equation}
where $q$ is any positive integer number and $c_k$ are some real
phenomenological coefficients. Obviously, the condition  $\hat N_i
[\bi u] = \hat N_i [u_i]$, is fulfilled: this perturbation is
local. After the regularisation the FP-image of this operator
appears as a sum, $\opn = \sum_{i=1}^N \opn_i$, with $N$ being the
number of samples and $\opn_i$ given by:
\begin{equation}\label{N1}
\opn_i = - \sum_{k=1}^q c_k \, r_i^{2k+1} \frac{\partial}{\partial
\phi_i}.
\end{equation}
Up to the first order in $\varepsilon$ the solution of
\eref{FP1N-pert} is then given by Eq.(\ref{1n1}), where one should
substitute in Eq.\eref{greenpert} the Green function of the
corresponding dimensionality, Eq.\eref{green}: $\mc G_M \to \mc
G_1$. Any local perturbation can be handled in the same way.

\subsection{More general systems: two and more zero-dispersion NLSEs coupled via perturbations}
\label{sec:pertub-local:mult} Now we proceed to more complicated
systems, e.g. when we need to consider a set of equations
(\ref{nl}) with coupling perturbations. In this subsection the
number of dimensions of the corresponding base systems is still
one, but the perturbation now cannot be treated as local. For
instance if we consider two nonlinear couplers or a single
birefringent fibre, then in place of Eq.(\ref{nl}) we obtain the
following system:
\begin{equation}\label{nl2}
\begin{split}
\frac{\partial u}{\partial
\zeta}&=\mathrm{i}|u|^2u+\eta^1(t,\zeta) + \varepsilon \hat
N^1[u(\zeta),v(\zeta)] \, ,
\\ \frac{\partial v}{\partial
\zeta} & =\mathrm{i}|v|^2v+\eta^2(t,\zeta) + \varepsilon \hat
N^2[u(\zeta),v(\zeta)] \, .
\end{split}
\end{equation}
Note that complex noises, $\eta^1$ and $\eta^2$, are the
independent WGNs, and may possibly have different intensities
$D_1$ and $D_2$. The perturbation operators, $\hat N^1[u,v]$ and
(or) $\hat N^2[u,v]$, are supposed to intermix the components $u$
and $v$ because otherwise the latter would be uncorrelated and one
could merely consider the first and the second equations of system
(\ref{nl2}) independently using the results of the previous
subsection.

It is convenient to introduce a second upper index to distinguish
the variables for $u$ (superscript ``1'') and $v$ (superscript
``2''), and opt for the following indexing rule: after the
regularisation we introduce vectors $\bi
r_{2N}=(r_1^1,\ldots,r_1^N,r_2^1, \ldots,r_2^N)$, and $\bphi_{2N}
= (\phi_1^1,\ldots, \phi_1^N,\phi_2^1,\ldots, \phi_2^N)$, where
\begin{equation}
\begin{split}\label{sampl2}
r_1^i &=|u(i\Delta t)|,\quad r_2^i=|v(i\Delta t)|, \quad
i=1,\ldots N \, ,
\\
 \phi_1^i = &\mathrm{Arg}[u(i\Delta t)], \quad
\phi_2^i=\mathrm{Arg}[v(i\Delta t)], \quad i=1,\ldots N \, .
\end{split}
\end{equation}
We seek the PDF of perturbed system, Eq.\eref{FP1N-pert}:
$\tilde{\mc P}_K(\bi r_K,\bphi_K|\,\bi r_K^0,\bphi^0_K;\zeta)$,
$K=2N$.  The perturbation in \eref{nl2} couples the dynamics (and
hence the statistics) of variables $(r_1^i,\phi_1^i)$ and
$(r_2^i,\phi_2^i)$ for $i=1,\ldots,N$. Since we have only two
(possibly) different noise intensities, $D_1$ and $D_2$, for this
case we can rewrite general expression \eref{greenpert} in more
simple explicit form:
\begin{multline}\label{greencouple}
 \bi{\mc{G}}_K(\zeta - \zeta'; \ \bphi_K - \bphi'_K; \bi r_K ,
\bi r'_K) = \prod_{i=1}^N \mc G_1^{D_1}(\zeta - \zeta'; \
\phi_1^i- \phi_1^{'i}; r_1^i, r_1^{'i}) \\ \times \mc
G_1^{D_2}(\zeta - \zeta'; \ \phi_2^i- \phi_2^{'i}; r_2^i,
r_2^{'i})\, ,
\end{multline}
where the superscripts of the propagators mean that one ought to
insert the corresponding noise intensity in the expression for
$\mc G_1$, Eq.\eref{green}. We can then employ the results of
subsection \ref{subsec:pert}.

The generalisation for the case when we have $q$ zero-dispersion
NLSEs coupled via the $q$-dimensional perturbation operator,
\[
\hat N  = \big( \hat N^1, \ldots, \hat N^q \big)^T ,
\]
is now straightforward. Firstly one should derive the FP-image
$\hat{\mc N}$ for the perturbation $\hat N$. Next, by analogy with
the case of two equations, one should introduce $K$-dimensional
vectors ($K = q \times N$): $\bi r_K =(r_1,\ldots,r_K)$ and
$\bphi_K = (\phi_1,\ldots,\phi_K)$. Again for the unperturbed
system we have $ P^K (\bi r_K,\,\bphi_K | \,\bi q_K^0,
\bphi_K^0;\zeta)$ in the product form \eref{factor}. To obtain the
expansion for function $\tilde{\mc P}^{K}$ we then use the
formulae given in subsection \ref{subsec:pert}. The expression for
the propagator can be given in a simplified form similar to
Eq.\eref{greencouple}: it is a product of $q$ equivalent
multipliers $\mc G_1^{D_k}$, $k=1, \ldots, q$, where the
corresponding noise intensities $D_k$ should be inserted in the
expression for the propagator Eq.\eref{green}.

\subsubsection{Coupling perturbations of uncoupled systems} Let us
now provide some examples of coupling perturbations.

\begin{itemize}
\item For birefringent fibres, with the coupling ($\sim
\varepsilon$) between the different polarisation components of
either linearly or circularly polarised light, we have (see e.g.
\cite{aa00}):
\begin{equation}\label{bf}
\hat N [u,v] = \left(\begin{array}{c}
\hat N_1[u,v] \\
\hat N_2[u,v] \end{array} \right) = \rmi \, \left(
\begin{array}{c}
A \,u |\, v|^2 + B \,v^2 \, u^* \\
A \,v |\, u|^2 + B \,u^2 \, v^* \end{array} \right) \, ,
\end{equation}
where $A$ and $B$ are the (real) constant coefficients of
cross-phase modulation and four-waves mixing. The corresponding
``polar'' FP-image of $\hat N$ in Eq. \eref{FP1N-pert} by virtue
of indexing scheme \eref{sampl2} can be presented as a sum
$\hat{\mc N} = \sum_{i=1}^{N} \hat{\mc N}_i$, where $N$ is, as
usual, the number of time-samples, and for $\hat{\mc N}_i$ we have
the following formula:
\begin{eqnarray}\label{biref}\nonumber
\opn_i &=& -\left( A + B \cos \big[ 2 (\phi_1^i - \phi_2^i)
\big]\right) \Biggl( (r_2^i)^2 \frac{\partial }{\partial \phi_1^i}
+ (r_1^i)^2 \frac{\partial}{\partial \phi_2^i} \Biggr) \\ &+& B
r_1^i \, r_2^i \sin \big[ 2(\phi_1^i-\phi_2^i)\big] \left( r_1^i
\frac{\partial}{\partial r_2^i} - r_2^i \frac{\partial}{\partial
r_1^i} \right) \, .
\end{eqnarray}

\item Now consider two nonlinear oscillators with small
eigenfrequencies and weak linear coupling (both terms $\sim
\varepsilon$). This is a special case of self-trapping model
\cite{kc86} in the presence of the additive noise. This system is
interesting in itself: it is integrable (in the noiseless case)
and possesses a finite degree of freedom analogue of soliton
solution (inhomogeneous state). Therefore this system can be
reckoned as one of the simplest model examples, where we can
investigate the action of noise on an integrable system. The
coupling operator $\hat N$ now is
\begin{equation}\label{rt}
\hat N [u,v] = \mathrm i \, \left( \begin{array}{c}
- \omega u  + \gamma (v - u) \\
- \omega v  + \gamma (u - v) \end{array} \right) \, ,
\end{equation}
where $\omega$ and $\gamma$ are the real parameters. The number of
independent variables is just four: $u \to \{ r_1, \phi_1\}$, $v
\to \{r_2,\phi_2\}$. The polar FP-image of $\hat N$ takes the form
\begin{eqnarray}\label{rots}\nonumber
\opn &=& (\omega + \gamma) \left( \frac{\partial}{\partial \phi_1}
+ \frac{\partial}{\partial \phi_2}\right) + \gamma \sin (\phi_2 -
\phi_1) \left( r_2 \frac{\partial}{\partial r_1} - r_1
\frac{\partial}{\partial r_2} \right) \\ &-& \gamma \cos (\phi_2 -
\phi_1) \left( \frac{r_2}{r_1} \, \frac{\partial}{\partial \phi_1}
+  \frac{r_1}{r_2} \, \frac{\partial}{\partial \phi_2} \right) \,
.
\end{eqnarray}
The expression for the propagator is very simple:
\begin{multline} \nonumber \bi{\mc{G}}_2(\zeta - \zeta'; \ \bphi - \bphi'; \bi r, \bi
r') \\ = \mc G_1^{D_1}(\zeta - \zeta'; \ \phi_1- \phi_1'; r_1,
r_1') \, \mc G_1^{D_2}(\zeta - \zeta'; \ \phi_2- \phi_2'; r_2,
r_2')\, ,
\end{multline}
where, as before, the expression for each $\mc G_1^{D_k}$ is given
by Eq.\eref{green}.

\item Consider a circular nonlinear fibre array comprising $q$
fibres with weak ($\sim \varepsilon$) linear coupling. Then
instead of (\ref{nl2}) we acquire a set of $q$ equations for $q$
functions $v_j$: $\bi{v}=(v_1, \ldots, v_q)$. The $q$-dimensional
coupling operator now becomes \cite{aa00}:
\begin{equation}\label{fa}
\hat N [\bi{v} ] = \left( \begin{array}{c}
\hat N_1[\bi{ v}] \\
\hat N_2[\bi{ v}] \\
\vdots \\
\hat N_q[\bi{ v}] \end{array} \right) = \rmi \, C \, \left(
\begin{array}{c}
 v_q  + v_2 \\
 v_1  + v_3 \\
 \vdots \\
 v_{q-1} + v_1
\end{array} \right) ,
\end{equation}
with $C$ a real constant. Again it is convenient to introduce the
double indexation scheme: the upper index will indicate the sample
number, the lower one corresponds to the field number: $v_1(i
\Delta t) \to \{ r_1^i, \phi_1^i\}, \ldots, v_q(i \Delta t) \to \{
r_q^i, \phi_q^i \}$. Then the image of the perturbation \eref{fa}
can be presented as a sum, $\opn = \sum_{i=1}^N \opn_i$ (with $N$
the number of samples), where
\begin{eqnarray} \nonumber \label{fan}
 \opn_i & =& C\sum_{n=1}^{q}\left[ A_n^i (\bi r_q^i, \bphi_q^i) \left( \cos \phi_n^i \,
\frac{\partial}{\partial r_n^i} -\frac{\sin \phi_n^i}{r_n^i} \,
\frac{\partial}{\partial \phi_n^i}\right)+ B_n^i(\bi r_q^i, \bphi_q^i) \right. \\
& \times &  \left. \left( \sin \phi_n^i \,
\frac{\partial}{\partial r_n^i} + \frac{\cos \phi_n^i}{r_n^i} \,
\frac{\partial}{\partial \phi_n^i} \right) \right],
\end{eqnarray}
and we used denotations
\[
\begin{split}
A_n^i (\bi r_q^i, \bphi_q^i) & = r_{n-1}^i \sin \phi_{n-1}^i +
r_{n+1}^i \sin
\phi_{n+1}^i \, ,\\
B_n(\bi r_q^i,\bphi_q^i) &= - r_{n-1}^i \cos \phi_{n-1}^i -
r_{n+1}^i \cos \phi_{n+1}^i \, .
\end{split}
\]
Here we have assumed that $(r_0^i,\phi_0^i) \equiv
(r_q^i,\phi_q^i)$ and
$(r_{q+1}^i,\phi_{q+1}^i)\equiv(r_1^i,\phi_1^i)$. The explicit
form for the propagator is now:
\[
 \bi{\mc{G}}_K(\zeta - \zeta'; \ \bphi_K - \bphi'_K; \bi r_K ,
\bi r'_K) = \prod_{i=1}^N \prod_{k=1}^q \mc G_1^{D_k}(\zeta -
\zeta'; \ \phi_k^i - \phi_k^{'i}; r_k^i, r_k^{'i})\, ,
\]
with $K = q \times N$.
\end{itemize}
\subsubsection{Birefringent fibre with weak four-wave mixing}
So far the original unperturbed system has always been equivalent
to the system of uncoupled oscillators, i.e. matrix $a_{ij}$ in
\eref{2rotn}, \eref{2rotfp} and \eref{greenN} has always been an
identity matrix and thus the dimensionality of the base system has
been one. This is not the case for a birefringent fibre with a
strong cross-phase modulation, i.e. if the inequality $A \gg
\varepsilon$ holds in Eq.\eref{bf}. Then one has to consider a
more general perturbed system at zero dispersion:
\begin{equation}\label{nl3}
\begin{split}
\frac{\partial u}{\partial
\zeta}&=\mathrm{i}(|u|^2+A|v|^2)u+\eta^1(t,\zeta) + \varepsilon
\hat N^1[u(\zeta),v(\zeta)] \, ,
\\ \frac{\partial v}{\partial
\zeta} & =\mathrm{i}(A|u|^2+|v|^2)v+\eta^2(t,\zeta) + \varepsilon
\hat N^2[u(\zeta),v(\zeta)] \, ,
\end{split}
\end{equation}
where $A$ is a real constant. Putting $A=0$ will correspond to
system \eref{nl2}. We see that now the dimensionality of the base
system is two, the matrix $a_{ij}$ of the base system has the
elements $a_{11}=a_{22}=1$, $a_{12}=a_{21} = A$. As a simplest
example of a perturbation for system \eref{nl3} we choose the
four-wave mixing (a special case of \eref{bf}):
\begin{equation}\label{bf1}
\hat N [u,v] = \rmi \, B \left( \begin{array}{c}
v^2 \, u^* \\
u^2 \, v^* \end{array} \right) ,
\end{equation}
with real constant $B$; the intensities of the noises, $\eta^1$
and $\eta^2$, can, of course, be different: $D_1$ and $D_2$.
Unlike the case of system \eref{bf}, now the perturbation
\eref{bf1} is a local one: $\hat N_i[\bi u, \bi v ] = \hat
N_i[u_i, v_i]$. The ``polar'' FP-image of $\hat N$ has been
already derived as a part of expression \eref{biref}, responsible
for the four-wave mixing, i.e. $\opn = \sum_{i=1}^{N} \opn_i$
with:
\begin{eqnarray}\label{Nbiref}\nonumber
\opn/B  &=& - \cos \big[ 2 (\phi_1^i - \phi_2^i) \big] \Biggl(
(r_2^i)^2 \frac{\partial }{\partial \phi_1^i} + (r_1^i)^2
\frac{\partial}{\partial \phi_2^i} \Biggr)  +  r_1^i \, r_2^i \sin
\big[ 2(\phi_1^i-\phi_2^i)\big] \\ &\times& \left( r_1^i
\frac{\partial}{\partial r_2^i} - r_2^i \frac{\partial}{\partial
r_1^i} \right) \, .
\end{eqnarray}
The propagator is given by
\begin{equation}
\label{greenpert1} \bi{\mc{G}}_K(\zeta - \zeta'; \ \bphi_K -
\bphi'_K; \bi r_K , \bi r'_K) = \prod_{i=1}^N \mc G_2(\zeta -
\zeta'; \ \bphi_2^i- \bphi_2^{' i};\bi r_2^i, \bi r^{' i}_2) \, ,
\end{equation}
with the corresponding noise intensities and coupling matrix
elements $a_{ij}$ inserted into general definition \eref{greenN}.

The cases of coupling (nonlocal) perturbations and the higher
number of dimensions of the corresponding base system can be
treated analogously.
\subsection{Discrete self-trapping equation}
It is instructive to give explicit expressions for noisy DSTE. In
this system the discreteness is postulated by the model itself,
the dimensionality of the base system is one, and we can readily
find the perturbation of FP operator applying the results above.
Now the perturbation reads (see equation \eref{dste}):
\begin{equation}\label{ndste}
\hat N[\bi u] = \rmi  \sum_{j=1}^N m_{nj} u_j \, ,
\end{equation}
where $N$ defines the number of discrete sites in the chain
(lattice). In this case the nonlocal perturbation \eref{ndste}
might couple all $N$ oscillators. Its FP-image is
\begin{equation}
\opn = \sum_{n,j=1}^{N} m_{nj} \, r_{j} \Biggl[ \sin \big(
\phi_{j} - \phi_n) \, \frac{\partial}{\partial r_n}\, - \,
\frac{\cos \big(\phi_{j}-\phi_n \big )}{r_n} \,
\frac{\partial}{\partial \phi_n}  \Biggr] ,
\end{equation}
and the propagator of the system factorises into a product of $N$
functions $\mc G_1^{D_i}$, with the corresponding noise
intensities.  We can then apply the general formulae from
subsection \ref{subsec:pert} to find the perturbed PDF.

\section{Second order dispersion as a perturbation}
\label{sec:nonlocal} In this  section we will consider more
complicated examples of nonlocal perturbations. We will
concentrate on one particular type of perturbations which is
nevertheless of great practical interest: the dispersive terms in
Kerr systems. Such terms can account for, for instance, a small
residual dispersion in a dispersion-managed fibre link. To start
we first study a scalar case of a single NLSE, and then proceed to
vector generalisations, i.e. to the system of coupled NLSEs.

\subsection{Scalar NLSE with weak second order
dispersion.} \label{sec:pertub-nonlocal}

In this subsection we will consider the perturbation in terms of
weak second order dispersion (SOD) in (\ref{nl}), with the
dimensionality of the base system being one:
\begin{equation}
\hat N[u]=\mathrm{i}\,\frac{d}{2}\,\frac{\partial^2 u}{\partial
t^2} . \label{sod}
\end{equation}
Here $d=\pm 1$ specifies the type of the dispersion (anomalous or
normal). For the sake of simplicity we consider the case $d=1$
since the choice of $d=-1$ only alters the sign in the appropriate
formulae. As we work with discretely sampled data we need to write
down a discrete analogue of the second derivative Eq.(\ref{sod}).
To do so we first define the forward finite difference as
$\boldsymbol{\delta} f_n = f_{n+1} -f_n$. The arbitrary precision
expansion for the 1D second derivative, $\partial^2 u(t)/\partial
\, t^2$, reads as \cite{a77}:
\begin{equation}\label{diff}
\frac{\partial^2 u}{\partial \, t^2} \Big|_{t=i\Delta t} =
\frac{1}{(\Delta t)^2} \ln^2(1 +\boldsymbol{\delta}) u_i
=\frac{1}{(\Delta t)^2} \sum_{n=2}^{\infty} b_n \,
\boldsymbol{\delta}^n u_i\, ,
\end{equation}
where we have applied regularisation procedure (\ref{regul}) and
$\boldsymbol{\delta}^n$ is the forward difference of the $n$th
order:
\[
\boldsymbol{\delta}^2 u_i = u_{i+1} - 2 u_i + u_{i-1}, \quad
\boldsymbol{\delta}^3 u_i = u_{i+2} -3 u_{i+1} + 3 u_i - u_{i-1}
\,, \quad {\rm etc}.
\]
The squared logarithm operator in representation (\ref{diff})
should be treated as a Taylor expansion in powers of
$\boldsymbol{\delta}$, and the expansion coefficients $b_n$ can be
easily calculated: $b_2 =1$, $b_3=-1$, $b_4=11/12$, etc.

Taking into account Eq.(\ref{diff}) we now write the discretised
perturbation $\hat{N}_i$ as:
\begin{equation}\label{ninf}
(\Delta t)^2 \hat N_i [\bi{ u}] = \mathrm i \ln^2(1 +
\boldsymbol{\delta}) u_i
 = \rmi \sum_{n=2}^{\infty} b_n \boldsymbol{\delta}^n u_i
\end{equation}
Further on it is convenient to use the explicit expression for the
difference operators in terms of binomial decomposition
\cite{a77}:
\[
\boldsymbol{\delta}^n u_i = \sum_{j=0}^{n}(-1)^j \binom{n}{j}
u_{i+n-j} \, .
\]
The Fokker-Planck image of the perturbation $\opn$ then becomes:
\begin{eqnarray}\label{ni}\nonumber
(\Delta t)^2 \opn &=& \sum_{i=1}^{N} \sum_{n=2}^{\infty}
\sum_{j=0}^{n}(-1)^j \binom{n}{j}  b_n \, r_{i+n-j} \Biggl[ \sin
\big( \phi_{i+n-j} - \phi_i) \,
\frac{\partial}{\partial r_i}\\
& - &\, \frac{\cos \big(\phi_{i+n-j}-\phi_i \big )}{r_i} \,
\frac{\partial}{\partial \phi_i}  \Biggr] .
\end{eqnarray}
Note that we write an infinite upper limit of summation over the
index $n$. However, of course, each difference is limited by the
total number of samples $N$, and all the variables outside the
sampling interval should be discarded. The form of the
perturbative expansion for the full PDF does not differ from that
given in previous sections for the non-dispersive perturbations.
The total propagator has the form of the $N$-product of functions
$\mc G_1$, each having the same noise intensity $D$.

The same procedure may be easily developed if one wishes to
account for a perturbative term in the form of a derivative of an
arbitrary order with either real or complex coefficient ($\sim
\varepsilon$). For example, the dispersion of arbitrary order
$\mathrm{i}^k \partial^k u/\partial t^k$ would result in the
change of the power of the logarithm operator in expressions
(\ref{diff}),(\ref{ninf}). Eventually one should arrive at the
same formulae with a different set of coefficients $b_n$, which,
in that case, have to be the Taylor coefficients for the expansion
of the corresponding power of the logarithm operator (one will
have to change the lower summation limit for $n$ from 2 to $k$ in
Eqs.(\ref{diff}-\ref{ni}) as well). It is worth mentioning that
system (\ref{nl1}) with the discrete term, given by (\ref{ninf})
but truncated at $n=2$, corresponds to the \textit{weakly discrete
NLSE} \cite{krb01} with additive noise. The increase of the number
of terms in expansion (\ref{ninf}) (and consequently in
(\ref{ni})) obviously increases the precision at the expense of
the computational time.

In the same manner we could redefine the dispersion operator via
the backward differences (see \cite{a77}) and obtain an expression
very similar to \eref{ni}. If we wish to preserve the
time-inversion symmetry, we might use the symmetrised form instead
of Eq.\eref{diff}: $\opn = (\opn_{forward} + \opn_{backward})/2$.

\subsection{Manakov equations with weak second order dispersion}
Next we develop a perturbation theory accounting for the weak
dispersion in vector systems, such as Manakov equations \eref{man}
(i.e. for the regularised system \eref{man1}). Now the number of
dimensions of the base system is two and we consider the nonlocal
perturbation of the form:
\begin{equation}\label{manpert}
\hat N [u,v] = \frac{\rmi}{2}\left( \begin{array}{c}
\frac{\partial^2 u}{\partial t^2} \\
\frac{\partial^2 v}{\partial t^2} \end{array}\right).
\end{equation}
We do not present here the explicit expressions for the PDF
expansion as they are very similar to those given in the previous
subsection, but outline only the general procedure. First we
choose the same regularisation we used in subsection
\ref{sec:pertub-local:mult} for two uncoupled equations. Full
coupling matrix $a_{ij}$ comprise a set of $2\times2$ blocks
placed on the main diagonal, with all elements within each block
equal to 1. The expression for the Fokker-Plank image of
perturbation \eref{manpert} is very similar to Eq.\eref{ni}. The
only difference is that now we again have to endow the variables
with an additional index to distinguish between the fields: $u_i
\to (r_1^i, \phi_1^i)$, $v_i \to (r_2^i, \phi_2^i)$. Now the
expression for the FP-image of perturbation \eref{manpert}
consists of the two summands, $\opn_1$ and $\opn_2$, each having
the form of Eq.\eref{ni}: the variables in Eq.\eref{ni} for
$\opn_1$ should bear additional index ``1'', and the same applies
to $\opn_2$. The propagator for the entire sampled system is a
direct product of $\mc G_2^{D_k}$, with the corresponding
(possibly different) noise intensities, $D_1$ and $D_2$, and the
elements of the coupling matrix for the base system,
$a_{11}=a_{12}=a_{21}=a_{22}=1$. All statements from the previous
subsection, concerning accounting for either the time inversion
symmetry or a higher order dispersion apply to this case as well.

{\em Remark on higher systems of coupled NLSEs.} Following the
preceding example one can build perturbation expressions for the
systems comprising more than two nonlinearly coupled NLSEs. The
number of coupled NLSEs, say $M$, simply defines the
dimensionality of the base oscillator system. If we have $M$
coupled fields, $v_1, \ldots, v_M$, then introducing an additional
index labelling the field number one finally arrives at the
expression for the FP-image of a dispersive perturbation in the
form of a sum, $\opn = \sum_k \opn_k$, where each $\opn_k$ has the
form of Eq.\eref{ni}. The propagator for the whole sampled system
is again a direct $N$-product of $\mc G_M^{D_k}$.

\section{Conclusion}
In the current paper we have proposed an approach for studying the
stochastic dynamics of a noise-driven systems with Kerr-type
nonlinearity. We presented the exact expression for the PDF (and
the propagator) for a system of nonlinearly coupled oscillators.
Aside from the study of nonlinear lattice models with the additive
WGN, we showed how to apply these results to the description of
signal propagation in a weakly nonlinear dispersive media driven
by WGN. We used the fact that in a system at zero dispersion the
PDF for the output signal (field) can be obtained analytically.
Using the propagators for a zero dispersion system we have built a
perturbation theory and obtained the corrections to the
unperturbed PDF in a variety of physically meaningful situations.
In particular we were able to obtain the corrections to the PDF in
the presence of the nonlocal perturbation, such as a second order
dispersion. The knowledge of the output signal statistics allows
one to calculate a multitude of important quantities like the
probability of error in the fibre optics communication, or, in
principle, Shannon capacity of the nonlinear communication
channel, modelled by Eq.(\ref{nl}).

Using the discrete time picture one can also examine the
statistics of more complex systems than those, described in the
current paper, like, for instance, a system of weakly coupled NLSE
with (also weak) linear or nonlinear nonlocal perturbations (e.g.
nonlinear dispersion). This would simply require increasing the
number of dimensions in the corresponding PDF. It is also possible
to consider the higher-dimensional noisy NLSE in the same fashion,
e.g. the 2+1 dimensional NLSE where the (discretised) weak second
order derivatives are reckoned as a perturbation.

We recognise that for the case of nonlocal perturbations, like
higher order dispersion, when the perturbation operator in the FPE
becomes quite involved, the corrections to the PDF can only be
calculated numerically. Let us recall that the traditional way of
calculating the output signal statistics numerically is to use
direct Monte Carlo simulations. The latter method, however has a
number of significant drawbacks. Normally the calculations are
quite time consuming and cannot be used for estimation of the
tails of the PDFs, something which is crucial for the performance
assessment in the fibre optical communications. Therefore the
perturbation technique developed here, with the iteration
procedure similar to \eref{+psi12} performed numerically, may
present an important method of choice for the systems where the
direct Monte Carlo approach fails. The proposed method relies only
on the approximate evaluation of integrals and infinite sums
without direct integration of the noisy equation. And finally, the
pertrubative approach devised in the present paper does not
require the noise to be small - which is usually implied in the
majority of the approximate methods in stochastic dynamics.
Therefore it can be used to analyse systems  far from the
equilibrium where the impact of noise on the signal degrees of
freedom becomes significant.

In conclusion we also point out that the exact analytical results
concerning the statistics of nonlinear coupled systems are still
few and far between. Thus we think that the exact general
expression for the propagator of coupled noisy nonlinear
oscillators \eref{greenN} (which coincides with a retarded Green
function), is quite important in itself and can serve as a
``reference point'' for the studies of similar systems.

\ack  Authors would like to thank Sergei Turitsyn and Igor
Yurkevich for the fruitful discussions and valuable comments. S.D.
would like to acknowledge the support from the Leverhulme Trust
Project F/00250/B. This work was also supported by NATO
Collaborative Linkage Program (No PST.CLG.980068).

\appendix
\section{Eigenfunctions of the Fokker-Planck operator}
\label{sec:appendix} \setcounter{section}{1} In this section we
calculate the eigenfunctions and eigenvalues of the FPE for single
oscillator \eref{FP1} and for system of nonlinearly coupled
oscillators \eref{FP1N}. We start from the former case. We wish to
solve the eigenvalue problem:
\begin{equation}
\opl_1 \Psi = s \Psi \label{eigen1} \, .
\end{equation}
Since the obvious periodical boundary conditions in $\phi$ it is
convenient to perform a Fourier transform with respect to $\phi$:
\begin{equation}
\label{ser-phi}
\begin{split}
\Psi(r,\Delta \phi)& = \sum_{\nu = -\infty}^{\infty} \mathrm{e}^{
\mathrm{i} \nu  \Delta \phi}
\Psi_\nu(r) \\
\Psi_\nu(r) & = \frac{1}{2\pi} \intop_{0}^{2 \pi}
\mathrm{e}^{-\mathrm{i} \nu \Delta \phi} \, \Psi(r,\Delta \phi)\,
\rmd \Delta \phi\, .
\end{split}
\end{equation}
After the transform the operator $\opl_1$ reads as
\begin{equation}\label{ml}
\opl_1^{\nu} = -D \left( \frac{1}{r} \frac{\partial}{\partial r}
\left( r \, \frac{\partial}{\partial r} \right) -
\frac{\nu^2}{r^2} \right) + \mathrm{i} \nu r^2 \, .
\end{equation}
To solve the eigenfunction equation we make the following
substitution:
\begin{equation}\label{ef}
\Psi_\nu(y) = y^{\, |\nu| /2} \exp(-\alpha_{\nu}y/2) \varphi(y)
\,,
\end{equation}
where $\alpha_{\nu} = (1+\mathrm{i}) \sqrt{\nu D/2}$ for $\nu>0$
and $\alpha_{\nu} = (1-\mathrm{i}) \sqrt{|\nu| D/2}$ for $\nu<0$,
$y = r^2/D$. After all the transformations Eq.\eref{eigen1} now
has the form:
\begin{equation}\label{eq}
y \frac{\rmd ^2 \varphi}{\rmd y^2} +(1+ |\nu| - \alpha_{\nu} y)
\frac{\rmd \varphi}{\rmd y} + z_{s \nu}\varphi =0,
\end{equation}
with $z_{s \nu} = (s - 2 \bigr[1 + |\nu|\bigl]\alpha_{\nu})/4$.
Equation (\ref{eq}) has the canonical form of the degenerated
hypergeometric equation \cite{Gr-Ryg}. Since we want our
eigenfunctions to decrease at infinity we must demand that $z_{s
\nu}/\alpha_\nu=n$, $n=0,1, \ldots $. This will determine the
discrete eigenvalues $s_{n\nu}$ of \eref{eigen1}, while the
corresponding eigenfunctions $\Psi_{n\nu}$ are expressed via
generalised Laguerre functions $\mm L^{\alpha}_{\beta}(\ldots)$ as
\cite{Gr-Ryg}:
\begin{equation}
\label{eigensystem1}
\begin{split}
& s_{n \nu}  =  2 \alpha_\nu \big( 2 n + 1 + |\nu|) \, , \\
& \Psi_{n\nu}(r,\phi)=\mm{e}^{\mm{i}\nu
\phi}\,\left(\frac{\alpha_\nu}{\pi
D}\right)^{1/2}\,\left(\frac{n!}{(n+|\nu|)!}\right)^{1/2}
\,z^{|\nu|/2} \,\exp[-z/2]\,\mm{L}_n^{|\nu|}(z), \\
&z=\alpha_\nu r^2/D \, .
\end{split}
\end{equation}
Operator $\mc L_1$ is of course not Hermitian. Therefore we need
to introduce a set of left-eigenfunctions $\tilde \Psi_{n\nu}$,
which are the eigenfunctions of the adjoint operator $\mc
L_1^\dag$ with eigenvalues $\tilde s_{n\nu}$:
\begin{equation}
\label{eigensystem2}
\begin{split}
& \tilde s_{n \nu}  = s^*_{n\nu} \, ,\\
& \tilde \Psi_{n\nu}(r,\phi)=\mm{e}^{\mm{i}\nu
\phi}\,\left(\frac{\alpha^*_\nu}{\pi
D}\right)^{1/2}\,\left(\frac{n!}{(n+|\nu|)!}\right)^{1/2}
\,z^{*|\nu|/2} \,\exp[-z^*/2]\,\mm{L}_n^{|\nu|}(z^*) \, .
\end{split}
\end{equation}
Eigenfunctions \eref{eigensystem1},\eref{eigensystem2} form a
biorthogonal system, i.e.
\begin{equation}
\intop_0^{2\pi} d\phi \intop^{\infty}_0 dr r \tilde
\Psi^*_{n\nu}(r,\phi) \Psi_{n'\nu'}(r,\phi) =
\delta_{nn'}\,\delta_{\nu \nu'} \, . \label{bi-orthogonal}
\end{equation}
Making use of the identity (see \cite{Gr-Ryg}):
\begin{equation}
\sum^{\infty}_{n=0} n! \,\frac{\mm L^{\alpha}_n (u)\, \mm
L^{\alpha}_n (v) \,
w^n}{\Gamma(n+\alpha+1)}=\frac{(uvw)^{-\alpha/2}}{1-w} \exp(-w
\frac{u+v}{1-w}) \, \mathrm{I}_{\alpha}\left(2
\frac{\sqrt{uvw}}{1-w}\right), \; |w|<1 \, ,\label{ident1}
\end{equation}
analytically extended to the complex plane, in the limit $w \to 1$
we can obtain the \textit{closure relation} for the left and right
eigenfunctions:
\begin{equation}
\sum_{n,\nu} \tilde \Psi^*_{n\nu}(r,\phi)\Psi_{n\nu}(r',\phi') =
r^{-1}\,\delta(r-r')\,\delta(\phi-\phi') \label{closure}
\end{equation}
(factor $r^{-1}$ in the r.h.s. is merely a jacobian).

In the same way, generalising the derivation of the elementary
basis \eref{eigensystem1}, \eref{eigensystem2}, we can built left-
and right-eigenfunction for the general $2M$-dimensional operator
$\opl_M$ in \eref{FP1N}. Using the Dirac notations one gains
\begin{equation} \label{eigensystem34}
\begin{split} |\, n_1, \ldots, n_M, \nu_1, \ldots, \nu_M \rangle & =
\frac{1}{\pi^{M/2}} \prod_{i=1}^{M} \rme^{\rmi \nu_i \phi_i}
\sqrt{\frac{\alpha_i}{D_i}}\left[
\frac{n_i!}{(n_i+|\nu_i|)!}\right]^{1/2}\\
&\times z_i^{|\nu_i|/2} \rme^{-z_i/2} \, \mm L_{n_i}^{|\nu_i|}(z_i) \, , \\
 \langle \, n_1, \ldots, n_M, \nu_1, \ldots, \nu_M | & =
\frac{1}{\pi^{M/2}} \prod_{i=1}^{M} \rme^{\rmi \nu_i \phi_i}
\sqrt{\frac{\alpha_i}{D_i}}\left[
\frac{n_i!}{(n_i+|\nu_i|)!}\right]^{1/2}\\
&\times z_i^{*|\nu_i|/2} \rme^{-z^*_i/2} \,
\mm L_{n_i}^{|\nu_i|}(z^*_i) \, , \\
z_i&=\alpha_i r_i^2/D \, .
\end{split}
\end{equation}
Here the quantities $\alpha_i(\nu_1, \ldots, \nu_M)$ are defined
as:
\begin{equation}
\label{cases} \alpha_i(\nu_1, \ldots, \nu_M) = \left\{
\begin{array}{ccc}
(1 + \rmi) \left(\frac{D_i}{2} \sum_{j=1 }^M a_{j\,i} \nu_j
\right)^{1/2}
& \text{for} &  \sum_{j=1 }^M a_{ji} \nu_j > 0 \, , \\
(1 - \rmi) \left(\frac{D_i}{2} \left|\sum_{j=1 }^M a_{ji} \nu_j
\right|\right)^{1/2} & \text{for} & \sum_{j=1 }^M a_{j\,i} \nu_j<0
\, .
\end{array}
\right.
\end{equation}
The corresponding (right) eigenvalues of operator $\opl_M$ are
\begin{equation} \label{eigen2}
 \, s_{n_1, \ldots, n_M \nu_1, \ldots, \nu_M} = 2  \sum_{i=1}^{M} \alpha_i\big( 2 n_i
+ 1 + |\nu_i|).
\end{equation}

Since we know the explicit expressions for eigenvalues, we know
the damping eigenrates ($\gamma = |\mathrm{Re}[s]|$) for each
eigenmode of FPE. These may be important for the investigation of
the long scale (large values of $\zeta$) evolution of the system
statistics, because for the case $\zeta \to \infty$ the statistics
of the system is determined by a single eigenmode with the lowest
value of $\gamma$. We can see that the damping rate for each
eigenmode is proportional to $\gamma(\nu ) \sim D^{1/2} \nu^{3/2}$
for a single oscillator or, for a particular realisation $(\nu_1,
\ldots, \nu_N)$, to $\gamma(\nu_1, \ldots, \nu_N) \sim
\sum_{i,j=1}^N D_i^{1/2} \nu_i (a_{j \, i}\nu_{j})^{1/2}$ for a
system of nonlinearly coupled oscillators. Note that the real
parts of eigenvalues \eref{eigensystem1},\eref{eigen2} are always
positive which means that equations \eref{FP1}, \eref{FP1N} do not
possess a stationary solution.


\begin{thebibliography}{}
\bibitem{e90} J.C. Eilbeck, Introduction to the discrete self-trapping
equation, in P.L. Christiansen and A.C. Scott eds., Davydov's
Solitons Revisited (Plenum Press, 1990) 473-483.

\bibitem{Elbeck85} J.C. Eilbeck, P.S. Lomdahl, and A.C. Scott, The discrete Self-trapping
equation, Physica D 16 (1985) 318-338.

\bibitem{krb01} P.G. Kevrekidis, K. \O. Rasmussen, and A. R.
Bishop, The discrete nonlinear Schrodinger equation: A survey of
recent results, Int. J. Mod. Phys. B 15 (2001) 2833-2900.

\bibitem{Elbeckonline} J.C. Eilbeck and  M. Johansson, The Discrete Nonlinear Schrodinger
equation - 20 Years on, Proc. of the 3rd Conf. Localization and
Energy Transfer in Nonlinear Systems, ed L V\'azquez et al, World
Scientific, New Jersey, 2003, 44-67.

\bibitem{cfk03} D.K. Campbell, S. Flach  and Yu.S. Kivshar,
Localizing energy through nonlinearity and discreteness, Phys.
Today 57 (2004) 43-49.

\bibitem{rcj98} K.\O. Rasmussen, P.L. Christiansen, M. Johansson, Yu.B. Gaididei, and
S.F. Mingaleev, Localized excitations in discrete nonlinear
Schrodinger systems: Effects of nonlocal dispersive interactions
and noise, Physica D 113 (1998) 134-151.

\bibitem{mt94} M.I. Molina, and G.P. Tsironis, Absence of localization in a nonlinear  random binary
alloy, Phys. Rev. Lett. 73 (1994) 464-467.

\bibitem{rck00} K.\O. Rasmussen, T. Cretegny, P.G. Kevrekidis, and
N. Gronbeck-Jensen, Statistical mechanics of a discrete nonlinear
system, Phys. Rev. Lett. 61 (2000) 3740-3743.

\bibitem{f02} T.D. Frank, Stability analysis of mean feld models described
by Fokker–Planck equations, Ann. Phys. (Leipzig) 11 (2002)
707-716.

\bibitem{Mec1} A.Mecozzi, Long-distance transmission at zero dispersion - combined effect of the  Kerr
nonlinearity and the noise of the in-line amplifiers, JOSA B 11
(1994) 462-469.

\bibitem{Mec2}  A.~Mecozzi, Limits to long-haul coherent transmission set by the Kerr nonlinearity and noise
of the in-line amplifierds, J. Lightwave Technol. 12 (1994)
1993-2000.


\bibitem{Mecozzi}  E.~Iannone, F.~Matera, A.~Mecozzi, and M.~Settembre,
Nonlinear Optical Communication Networks (John Wiley \& Sons,
1998).

\bibitem{Agraval} G.P. Agrawal, Nonlinear fiber optics,
(Academic Press, San Diego, 2001).

\bibitem{aa00} Akhmediev N.N., Ankiewich A, Solitons.
Nonlinear pulses and beams, (Chapman \& Hall, 2000).

\bibitem{Hausrev} H.A.~Haus, W.S.~Wong, Solitons in optical
communications, Rev. Mod. Phys. 68 (1996) 423-444.

\bibitem{fklt}  G.~Falkovich, I.~Kolokolov, V.~Lebedev, V. Mezentsev, and
S.~Turitsyn, Non-Gaussian error probability in optical soliton
transmission, Physica D 195 (2004) 1--28.

\bibitem{OSA03} S.A.~Derevyanko, S.K.~Turitsyn, and D.A. Yakushev,
Non-Gaussian statistics of an optical soliton in the presence of
amplified spontaneous emission, Opt. Lett. 28 (2003) 2097 - 2099.


\bibitem{lk97} T.I. Lakoba, D.J. Kaup, Perturbation theory for the Manakov soliton and its applications
to pulse propagation in randomly birefringent fibers, Phys. Rev. E
56 (1997) 6147-6165.

\bibitem{kl01} T. Kanna and M. Lakshmanan, Exact soliton solutions, shape changing collisions,
and partially coherent solitons in coupled nonlinear Schrodinger
equations, Phys. Rev. Lett. 86 (2001) 5043-5046.

\bibitem{s84} A. C. Scott, Launching a Davydov soliton 1. Soliton analysis, Phys. Scr. 29 (1984) 279-283.

\bibitem{cas95} S. Chakravarty, M. J. Ablowitz, J. R. Sauer, and R. B.
Jenkins, Multisoliton interactions and wavelength-division
multiplexing, Opt. Lett. 20 (1995) 136-138.

\bibitem{yb98} C. Yeh and L. Bergman, Enhanced pulse compression in a nonlinear fiber by a
wavelength division multiplexed optical pulse, Phys. Rev. E 57
(1998) 2398-2404.

\bibitem{Risken}  H.~Risken, The Fokker-Planck Equation, (Springer,
1996).

\bibitem{Tur00} S.K.~Turitsyn, E.G.~Turitsyna, S.B.~Medvedev, and
M.P.~Fedoruk, Averaged model and integrable limits in nonlinear
double-periodic Hamiltonian systems, Phys. Rev. E 61 (2000)
3127-3132.

\bibitem{PRL03} K.S.~Turitsyn, S.A.~Derevyanko,
I.V.~Yurkevich, and S.K.~Turitsyn, Information capacity of optical
fiber channels with zero average dispersion, Phys. Rev. Lett. 91
(2003) art. no. 203901.

\bibitem{kc86} V.M. Kenkre, and D.K. Campbell, Self-trapping on a dimer - time-dependent solutions of
a discrete nonlinear Shrodinger equation, Phys. Rev. B.  34 (1986)
4959-4961.

\bibitem{a77} W.F. Ames, Numerical Methods for  Partial Differential
Equations, (Academic Press, 1977).


\bibitem{Gr-Ryg} I.S. Gradshteyn, I.M. Ryzhik, Alan Jeffrey (Editor),
Table of Integrals, Series and Products, (Academic Press, New
York, 2002).

\end{thebibliography}
\end{document}